\documentclass[%
 aip,
 amsmath,
 amssymb,
 reprint,%
]{revtex4-1}

\usepackage{amsfonts}
\usepackage{amssymb}
\usepackage{graphicx}
\usepackage{amsmath}
\usepackage{bm}
\DeclareMathAlphabet{\mathcal}{OMS}{cmsy}{m}{n}
\usepackage[english]{babel}
\usepackage{color}
\usepackage[version=3]{mhchem} 

\definecolor{darkgreen}{RGB}{40,110,5}

\usepackage{ulem}  
\usepackage{dcolumn}

\usepackage{amsfonts}
\usepackage{amssymb}
\usepackage{graphicx}
\usepackage{amsmath}
\usepackage{bm}
\DeclareMathAlphabet{\mathcal}{OMS}{cmsy}{m}{n}
\usepackage[english]{babel}
\usepackage{color}
\usepackage[version=3]{mhchem} 

\definecolor{darkgreen}{RGB}{40,110,5}

\usepackage{ulem}  
\usepackage{dcolumn}

\usepackage[utf8]{inputenc}
\usepackage[T1]{fontenc}
\usepackage{mathptmx}
\usepackage{etoolbox}

\makeatletter
\def\@email#1#2{%
 \endgroup
 \patchcmd{\titleblock@produce}
  {\frontmatter@RRAPformat}
  {\frontmatter@RRAPformat{\produce@RRAP{*#1\href{mailto:#2}{#2}}}\frontmatter@RRAPformat}
  {}{}
}%
\makeatother
\begin{document}

\title{Dynamics of the spin-boson model: the effect of bath initial conditions}
\author{Lipeng Chen*}
\affiliation{Zhejiang Laboratory, Hangzhou 311100, China}
\email{chenlp@zhejianglab.com}
\author{Yiying Yan}
\affiliation{Department of Physics, School of Science, Zhejiang University of Science and Technology, Hangzhou 310023, China}
\author{Maxim F. Gelin*}
\affiliation{School of Science, Hangzhou Dianzi University, Hangzhou 310018, China}
\email{maxim@hdu.edu.cn}
\author{Zhiguo Lü*}
\affiliation{Key Laboratory of Artificial Structures and Quantum Control (Ministry of Education), School of Physics and Astronomy, Shanghai Jiao Tong University, Shanghai 200240, China}
\email{zglv@sjtu.edu.cn}

\begin{abstract}
Dynamics of the (sub-)Ohmic spin-boson model under various bath initial conditions is investigated by employing the Dirac-Frenkel time-dependent variational approach with the multiple Davydov $\mathrm{D_1}$ ansatz in the interaction picture. The validity of our approach is carefully checked by comparing results with those of the hierarchy equations of motion method. By analyzing the features of nonequilibrium dynamics, we identify the phase diagrams for different bath initial conditions. We find that for spectral exponent $s<s_c$, there exists a transition from coherent to quasicoherent dynamics with increasing the coupling strengths. For $s_c<s\leq{1}$, the coherent to incoherent crossover occurs at a certain coupling strength, and the quasicoherent dynamics emerges at much larger couplings. The initial preparation of the bath has considerable influence on the dynamics.  
\end{abstract}

\maketitle

\section{Introduction}

Decoding the intricate interplay between a quantum system and its environment is a long-term pursuit in the study of static and dynamical properties of open quantum systems. \cite{Leggett,Weiss} A paradigmatic model to describe the effect of the dissipative environment on a quantum system is the so-called spin-boson model (SBM), where a two-level system is coupled to a bath consisting of infinite number of harmonic oscillators. The SBM has found widespread applications ranging from quantum computation, \cite{SBMQC1,SBMQC2} superconducting qubit architectures, \cite{SBMSQ1,SBMSQ2} nanomechanical devices \cite{SBMND1,SBMND2} to the electron transfer in biological molecules. \cite{SBMET1,SBMET2}   

The effect of a dissipative environment on the quantum system is specified by the so-called spectral density $J(\omega)$, which has a typical form of $J(\omega)\sim\omega^s$ with spectral exponent $s$. Depending on the value of $s$, the bath is classified as Ohmic ($s=1$), sub-Ohmic ($s<1$), and super-Ohmic ($s>1$). For the Ohmic case, it is well established that there exists a coherent-incoherent dynamic transition, as well as a quantum phase transition from a delocalized phase to a localized one when increasing the system-bath interaction. Compared to the Ohmic case, the dominance of low-frequency bath modes in the sub-Ohmic SBM leads to peculiar static and dynamic properties. In recent years, advanced numerical techniques have been developed to treat the sub-Ohmic SBM. By using a generalization of Wilson's numerical renormalizaiton group, Bulla et al. obtained a line of continuous delocalized-localized phase transition for  $0<s<1$. \cite{Bulla2003} The application of a continuous time cluster Monte Carlo algorithm to the sub-Ohmic SBM has shown that the critical exponents are classical, mean-field like for $s<1/2$. \cite{Bulla2009} A numerical technique based on a sparse polynomial space representation has been utilized to study the quantum phase transition of the sub-Ohmic SBM, yielding a phase diagram in perfect agreement with the quantum Monte Carlo method. \cite{Fehske} 

Due to strong non-Markovian effects induced by the low-frequency bath modes, it is notoriously difficult to accurately simulate quantum dynamics of the sub-Ohmic SBM. Using the multilayer multiconfiguration time-dependent Hartree (ML-MCTDH) method, it is shown that for $s\leq{1}$ the weakly damped coherent dynamics becomes gradually overdamped with increasing system-bath coupling strengths, followed by the delocalization-localization transition for even larger coupling strengths. \cite{WangH1,WangH2} For an initially polarized heat bath, it is found that low-frequency bath modes can generate an effective time dependent bias for an intrinsically unbiased two level system, which in turn is closely connected to the dynamic crossover from damped oscillatory to incoherent dynamics. \cite{PNalbach} Numerical simulations based on the path integral Monte Carlo method have further demonstrated that nonequilibrium coherent dynamics exists even under ultrastrong dissipation for $0<s<1/2$. \cite{KastPRL,KastPRB} The dynamic phase diagram with a critical spectral exponent $s_c=0.4$ was determined by the multiple Davydov $\mathrm{D_1}$ ansatz, where one observes no overdamping under arbitrary system-bath interaciton for $s<s_c$, and the coherent-incoherent crossover and a reemergence of the coherent state at very large couplings for $s_c<s\leq{1}$. \cite{WangLu} Quite recently, by employing a time-evolving matrix product operator approach, Otterpohl et al. identified a novel pseudocoherent phase beside the well-known coherent and incoherent phases for an initially factorized bath. \cite{OtterpohlPRL} This phase is characterized by the aperiodic dynamics with a single oscillatory minimum, and the dynamics turns incoherent for infinite bath cutoff frequency. 

In this paper, we proposed a new approach based on the multiple Davydov $\mathrm{D_1}$ ansatz, which is capable of accurately simulating zero-temperature quantum dynamics of the (sub-)Ohmic SBM under various initial preparations of the bath. To this end, the original Hamiltonian is transformed to a time-dependent Hamiltonian in the interaction picture with the help of the displaced oscillator state for each bath mode, and the resulting driven dynamics is easily handled by the multiple Davydov $\mathrm{D_1}$ ansatz. The reliability of the new approach is carefully checked by comparisons with the hierarchy equations of motion (HEOM) method. We systematically studied the dynamics of the (sub-)Ohmic SBM over a broad range of parameters, i.e., system-bath coupling strengths, spectral exponent $s$ and the bath initial conditions, and presented a complete picture for the zero-temperature dynamic phase diagram of the (sub-)Ohmic SBM. 

The paper is organized as follows. In Sec II, we introduced the model and the new approach based on the multiple Davydov $\mathrm{D_1}$ ansatz. In Sec III, the dynamics of the (sub-)Ohmic SBM is systematically investigated, and the dynamic phase diagrams under various bath initial conditions are sketched. Finally, conclusions are drawn in Sec IV.

\section{Theory}
\subsection{Hamiltonian}
The Hamiltonian of the SBM can be written as (throughout this paper, we set $\hbar=1$) 
\begin{equation}
\hat{H}=\hat{H}_{\mathrm{S}}+\hat{H}_{\mathrm{B}}+\hat{H}_{\mathrm{SB}}
\end{equation}
where 
\begin{eqnarray}
\hat{H}_{\mathrm{S}}&=&-\frac{1}{2}\Delta\hat{\sigma}_x+\frac{1}{2}\epsilon\hat{\sigma}_z\\
\nonumber\\ 
\hat{H}_{\mathrm{B}}&=&\sum_k\omega_k\hat{b}_k^{\dagger}\hat{b}_k \nonumber \\
\hat{H}_{\mathrm{SB}}&=&\frac{1}{2}\hat{\sigma}_z\sum_k\lambda_k(\hat{b}_k^{\dagger}+\hat{b}_k)
\end{eqnarray}
Here, $\hat{\sigma}_x$ and $\hat{\sigma}_z$ are Pauli matrices, and $\hat{b}_k^{\dagger}$($\hat{b}_k$) is the creation (annihilation) operator of the $k$th bosonic mode with frequency $\omega_k$. $\epsilon$ and $\Delta$ are spin bias and bare tunneling constant, respectively. $\lambda_k$ is the coupling strength between the spin and the $k$th bath mode, which can be characterized by a spectral density 
\begin{equation}\label{SD}
J(\omega)=\sum_k|\lambda_k|^2\delta(\omega-\omega_k)=2\alpha\omega_c^{1-s}\omega^se^{-\omega/\omega_c}
\end{equation}
where $\omega_c$ is the cutoff frequency, $\alpha$ is the dimensionless coupling strength, and $s$ is the spectral exponent. In this paper, we focus on the sub-Ohmic bath due to the inherent difficulty of treating strongly non-Markovian dynamics induced by low frequency bath modes. \cite{LZGPRB,MPSPRB}  

\subsection{The shifted bath initial condition}
It has been shown that the initial preparation of the bath has considerable influence on the dynamics of the sub-Ohmic SBM. \cite{PNalbach,KastPRL,WangLu} Two kinds of bath initial conditions are often considered. One is the factorized initial condition where the spin is in thermal equilibrium with the bath at temperature $T$ and at initial time $t=0$, i.e.,  
\begin{equation}
\hat{\rho}(t=0)=\hat{\rho}_{\mathrm{S}}(t=0)\otimes\hat{\rho}_{\mathrm{B}}
\end{equation}
where $\hat{\rho}(t=0)$ and $\hat{\rho}_{\mathrm{S}}(t=0)$ are the initial total and system density matrices, respectively, and $\hat{\rho}_{\mathrm{B}}=e^{-\beta\hat{H}_{\mathrm{B}}}/\mathrm{Tr}{e^{-\beta\hat{H}_{\mathrm{B}}}}$ with inverse temperature $\beta=1/T$. The other is the shifted initial condition, $\hat{\rho}(t=0)=\hat{\rho}_{\mathrm{S}}(t=0)\otimes\hat{\rho}_{\mathrm{B\mu}}$, with $\hat{\rho}_{\mathrm{B\mu}}=e^{-\beta(\hat{H}_{\mathrm{B}}+\hat{H}_{\mathrm{SB}}|_{\hat{\sigma}_z=\mu})}/\mathrm{Tr}e^{-\beta(\hat{H}_{\mathrm{B}}+\hat{H}_{\mathrm{SB}}|_{\hat{\sigma}_z=\mu})}$. Here, $\mu$ is a parameter characterizing the polarization of the bath. $\mu=1$ and $\mu=0$ correspond to the polarized and factorized initial conditions, respectively. For $0<\mu<1$, the bath is prepared between these two limits.

At zero temperature, the shifted initial condition corresponds to the case where the bath is prepared in the ground state of the Hamiltonian $\sum_k\omega_k\hat{b}_k^{\dagger}\hat{b}_k+\frac{\hat{\sigma}_z}{2}\sum_k\lambda_k(\hat{b}_k^{\dagger}+\hat{b}_k)|_{\hat{\sigma}_z=\mu}$. Therefore, $\hat{\rho}_{\mathrm{B\mu}}$ can be represented with the help of a displaced-oscillator state for each mode, i.e.,  
\begin{equation}
|y_k\rangle=\hat{\mathrm{D}}(y_k)|0_k\rangle=\exp\left(-\mu\frac{\lambda_k}{2\omega_k}(\hat{b}_k^{\dagger}-\hat{b}_k)\right)|0_k\rangle
\end{equation}
as $\hat{\rho}_{\mathrm{B\mu}}=\hat{\mathrm{D}}(\mathbf{y})|\mathbf{0}\rangle\langle\mathbf{0}|\hat{\mathrm{D}}^{\dagger}(\mathbf{y})$. Here $\hat{\mathrm{D}}(\mathbf{y})=\prod_k\hat{\mathrm{D}}(y_k)$ and $|\mathbf{0}\rangle=\prod_k|0_k\rangle$. The total density matrix can be obtained as 
\begin{eqnarray}\label{rhot}
\hat{\rho}(t)&=&e^{-i\hat{H}_{\mathrm{B}}t}\hat{U}_I(t)\hat{\rho}_{\mathrm{S}}(0)\hat{\rho}_{\mathrm{B\mu}}\hat{U}_I^{\dagger}(t)e^{i\hat{H}_{\mathrm{B}}t} \nonumber \\
&=&e^{-i\hat{H}_{\mathrm{B}}t}\hat{U}_I(t)\hat{\rho}_{\mathrm{S}}(0)\hat{\mathrm{D}}(\mathbf{y})|\mathbf{0}\rangle\langle\mathbf{0}|\hat{\mathrm{D}}^{\dagger}(\mathbf{y})\hat{U}_I^{\dagger}(t)e^{i\hat{H}_{\mathrm{B}}t}
\end{eqnarray}
where $\hat{U}_I(t)$ is the time evolution operator in the interaction picture with the corresponding Hamiltonian $\hat{H}(t)=e^{i\hat{H}_\mathrm{B}t}(\hat{H}_\mathrm{S}+\hat{H}_{\mathrm{SB}})e^{-i\hat{H}_\mathrm{B}t}=\hat{H}_{\mathrm{S}}+\frac{\hat{\sigma}_z}{2}\sum_k\lambda_k(\hat{b}_k^{\dagger}e^{i\omega_kt}+\hat{b}_ke^{-i\omega_kt})$. The expectation value of any operator $\hat{O}$ can be expressed as 
\begin{eqnarray}
\begin{aligned}
\langle\hat{O}(t)\rangle=&\mathrm{Tr}\left[\hat{O}\hat{\rho}(t)\right]   \\
=&\mathrm{Tr}\bigg[\hat{\tilde{O}}_I(t)\hat{\tilde{U}}_I(t)\hat{\rho}_{\mathrm{S}}(0)|\mathbf{0}\rangle\langle\mathbf{0}|\hat{\tilde{U}}_I^{\dagger}(t)\bigg]
\end{aligned}
\end{eqnarray}
where we have introduced the transformed operator $\hat{\tilde{O}}_I(t)$ in the interaction picture 
\begin{equation}
\hat{\tilde{O}}_I(t)=\hat{\mathrm{D}}^{\dagger}(\mathbf{y})e^{i\hat{H}_{\mathrm{B}}t}\hat{O}e^{-i\hat{H}_{\mathrm{B}}t}\hat{\mathrm{D}}(\mathbf{y})
\end{equation}
and the transformed time evolution operator   $\hat{\tilde{U}}_I(t)$ 
\begin{equation}\label{TTE}
\hat{\tilde{U}}_I(t)=\hat{\mathrm{D}}^{\dagger}(\mathbf{y})\hat{U}_I(t)\hat{\mathrm{D}}(\mathbf{y})
\end{equation}
with the corresponding transformed Hamiltonian $\hat{\tilde{H}}(t)$
\begin{eqnarray}\label{THt}
\hat{\tilde{H}}(t)&=&\hat{\mathrm{D}}^{\dagger}(\mathbf{y})\hat{H}(t)\hat{\mathrm{D}}(\mathbf{y})  \nonumber \\
&=&\hat{H}_{\mathrm{S}}+\frac{\hat{\sigma}_z}{2}\left[y(t)+y^{*}(t)\right]+\frac{\hat{\sigma}_z}{2}\sum_k\lambda_k(\hat{b}_k^{\dagger}e^{i\omega_kt}+\hat{b}_ke^{-i\omega_kt}) \nonumber \\
\end{eqnarray}
Here, $y(t)=\sum_k\lambda_ky_ke^{-i\omega_kt}=\sum_k\lambda_k(-\mu\frac{\lambda_k}{2\omega_k})e^{-i\omega_kt}$, and $y(t)+y^{*}(t)=-\mu\int_0^{\infty}\frac{J(\omega)\cos(\omega{t})}{\omega}d\omega$. For the bath spectral density of Eq.~(\ref{SD}), we have
\begin{equation}\label{yt}
y(t)+y^{*}(t)=-2\mu\alpha\omega_c\cos[s\arctan(\omega_ct)]\Gamma(s)(\omega_c^2t^2+1)^{-s/2} 
\end{equation}
with $\Gamma(s)$ denoting Gamma function. 

\subsection{Time dependent variational approach with the multiple Davydov $\mathrm{D}_1$ ansatz}
The time evolution of the wave function $|\Psi(t)\rangle$ for the transformed Hamiltonian $\hat{\tilde{H}}(t)$ of Eq.~(\ref{THt}) satisfies the time-dependent Schr\"{o}dinger equation 
\begin{equation}\label{SET}
i\frac{\partial}{\partial{t}}|\Psi(t)\rangle=\hat{\tilde{H}}(t)|\Psi(t)\rangle
\end{equation}
In this work, we utilize the multiple Davydov $\mathrm{D}_1$ trial state, also known as the multi-$\mathrm{D}_1$ ansatz, to solve Eq.~(\ref{SET}). The time-dependent version of the multi-$\mathrm{D}_1$ ansatz \cite{Davydov1,Davydov2,Davydov3} reads 
\begin{eqnarray}\label{MD1}
\begin{aligned}
|\mathrm{D}_1^M(t)\rangle=&|+\rangle\sum_{n=1}^MA_n(t)\exp\left(\sum_kf_{nk}(t)\hat{b}_k^{\dagger}-\mathrm{H.c.}\right)|\mathbf{0}\rangle \\
&+|-\rangle\sum_{n=1}^MB_n(t)\exp\left(\sum_kg_{nk}(t)\hat{b}_k^{\dagger}-\mathrm{H.c.}\right)|\mathbf{0}\rangle
\end{aligned}
\end{eqnarray}
where $|\pm\rangle$ are the eigenstates of $\hat{\sigma}_z$ with eigenvalues $\pm{1}$, and $A_n(t)$ ($B_n(t)$) are the amplitudes in state $|+\rangle$ ($|-\rangle$). $f_{nk}(t)$ and $g_{nk}(t)$ are the phonon displacements with $n$ and $k$ denoting the $n$th coherent state and the $k$th bath mode. $\mathrm{H.c.}$ represents the Hermitian conjugate and $M$ is the multiplicity. 

The equations of motion for the variational parameters $A_n$, $B_n$, $f_{nk}$ and $g_{nk}$ can be obtained by the  time-dependent variational principle \cite{VPDF} via
\begin{equation}\label{VPEOM}
\langle\delta\mathrm{D}_1^M(t)|i\partial_t-\hat{\tilde{H}}(t)|\mathrm{D}_1^M(t)\rangle=0   
\end{equation}
where $\langle\delta\mathrm{D}_1^M(t)|$ is the variation of the multi-$\mathrm{D}_1$ ansatz. The readers are referred to Appendix \ref{AppSBMVP} for a detailed derivation of the equations of motion for $A_n$, $B_n$, $f_{nk}$ and $g_{nk}$.   

To perform numerical simulation, one needs to specify $\omega_k$ and $\lambda_k$. We follow a discretization procedure widely employed in the ML-MCTDH method to obtain $\omega_k$ and $\lambda_k$. $\omega_k$ and $\lambda_k$ can be obtained via the relation \cite{WangHMCTDHSD}
\begin{equation}
\lambda_k^2=\frac{J(\omega_k)}{\xi(\omega_k)}
\end{equation}
where $\xi(\omega)$ is a density of frequencies satisfying 
\begin{equation}
\int_0^{\omega_k}d\omega\xi(\omega)=k,\quad{k=1,\cdots,\mathrm{N}_b}
\end{equation}
and $\xi(\omega)$ is normally chosen as 
\begin{equation}
\xi(\omega)=\frac{\mathrm{N}_b+1}{\omega_c}e^{-\omega/\omega_c}
\end{equation}
with $\mathrm{N}_b$ being the number of bath modes.

\subsection{Observables}
In the SBM, the physical observables of interest are expectation values of $\hat{\sigma}_i$, $i=x,y,z$, which are defined as 
\begin{eqnarray}\label{Pxyz}
P_i(t)=&&\langle\hat{\sigma}_i(t)\rangle=\langle\mathrm{D}_1^M(t)|\hat{D}^{\dagger}(\mathbf{y})e^{i\hat{H}_{\mathrm{B}}t}\hat{\sigma}_ie^{-i\hat{H}_{\mathrm{B}}t}\hat{D}(\mathbf{y})|\mathrm{D}_1^M(t)\rangle \nonumber \\
=&&\langle\mathrm{D}_1^M(t)|\hat{\sigma}_i|\mathrm{D}_1^M(t)\rangle,i=x,y,z
\end{eqnarray}
Here, $P_x(t)$ and $P_y(t)$ are real and imaginary components of the coherence, while $P_z(t)$ represents the population difference. By substituting Eq.~(\ref{MD1}) into Eq.~(\ref{Pxyz}), one obtains 
\begin{eqnarray}
P_x(t)=&&\sum_{ln}^M\left[A_l^{*}B_nR_{ln}^{(f,g)}+B_l^{*}A_nR_{ln}^{(g,f)}\right]  \\
P_y(t)=&&-i\sum_{ln}^M\left[A_l^{*}B_nR_{ln}^{(f,g)}-B_l^{*}A_nR_{ln}^{(g,f)}\right] \\
P_z(t)=&&\sum_{ln}^M\left[A_l^{*}A_nR_{ln}^{(f,f)}-B_l^{*}B_nR_{ln}^{(g,g)}\right]
\end{eqnarray}
where $R_{ln}^{(f,f)}$, $R_{ln}^{(g,g)}$, $R_{ln}^{(f,g)}$ and $R_{ln}^{(g,f)}$ are defined in Appendix \ref{AppSBMVP}. 

\section{Results and discussions}
\subsection{Numerical comparisons with the HEOM method}
\begin{figure}[]
\centering
\includegraphics[width=8.5cm]{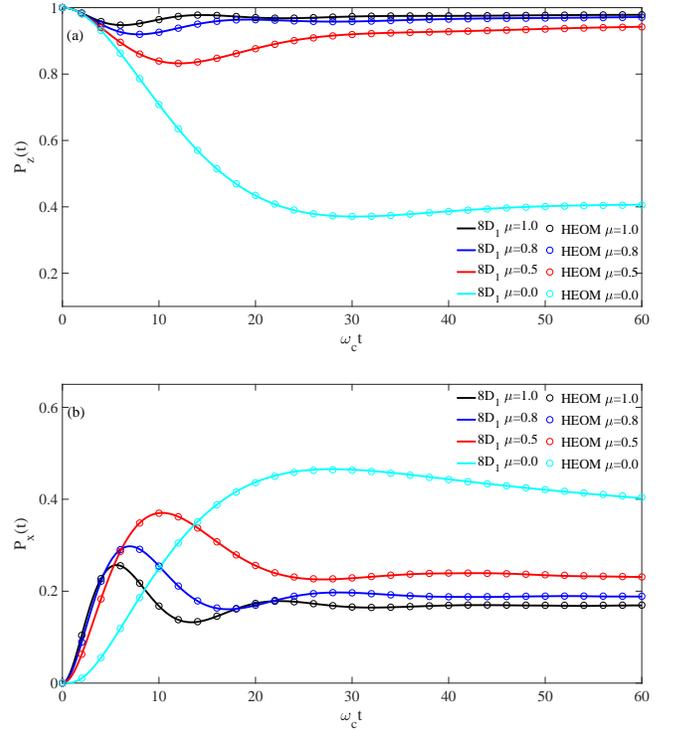}
\caption{Dynamics of $P_z(t)$ (a) and $P_x(t)$ (b) obtained by the multi-$\mathrm{D_1}$ ansatz with $M=8$ and the HEOM method for different initial bath preparation parameters, $\mu=1$, $0.8$, $0.5$, $0$. Other parameters are $s=0.25$, $\alpha=0.1$, $\Delta/\omega_c=0.1$, $\epsilon=0$, $\mathrm{N_b}=200$, and $T=0$.
}
\label{fig:s025diffu}
\end{figure}

We first check the numerical reliability of the proposed approach by comparisons with the HEOM method. By fitting the bath correlation function as a sum of exponentials and adopting an effective time-dependent Hamiltonian, we developed a new HEOM method which can calculate the dynamics of the SBM at zero temperature and under various bath initial conditions (see Appendix \ref{HEOMSBM} for details). Figure~\ref{fig:s025u1} plots the dynamics of $P_z(t)$ and $P_x(t)$ as calculated by the multi-$\mathrm{D_1}$ ansatz with $M=8$ and the HEOM method for different coupling strengths. We consider a polarized initial condition for the bath ($\mu=1$), and the other parameters are $s=0.25$, $\Delta/\omega_c=0.1$, $\epsilon=0$, $\mathrm{N_b}=200$, and $T=0$. For all coupling strengths, $P_z(t)$ and $P_x(t)$ obtained from the multi-$\mathrm{D_1}$ ansatz are in perfect agreement with those calculated by the HEOM method. It is found that the oscillation frequencies for both $P_z(t)$ and $P_x(t)$ increase with increasing system-bath coupling, and the oscillatory behavior survives even under ultrastrong coupling of  $\alpha=0.3$. As discussed in Refs.~\cite{PNalbach,KastPRL}, the polarized bath induces an effective time-dependent bias $\epsilon_p(t)$ (see Eq.~(\ref{timebias})), which leads to the fast decay to the quasi-equilibrium and ultraslow dynamics of the quasi-equilibrium itself. The persistence of coherent dynamics is also strongly influenced by the initial preparations of the bath, as illustrated in Fig.~\ref{fig:s025diffu}. One can readily observe that the exact dynamics of $P_z(t)$ and $P_x(t)$ for different initial bath preparation parameters, $\mu=1$, 0.8, 0.5, 0, are again well reproduced by the multi-$\mathrm{D}_1$ ansatz with $M=8$. The oscillation frequencies for both $P_z(t)$ and $P_x(t)$ are found to decrease with decreasing $\mu$. It should be noted that different initial preparations of the bath can be equivalently described by an effective time-dependent bias $\epsilon_p(t)$, which is proportional to $\mu$ via the $y(t)$ term (see Eq.~(\ref{yt})). 
  
\begin{figure}[]
\centering
\includegraphics[width=8.5cm]{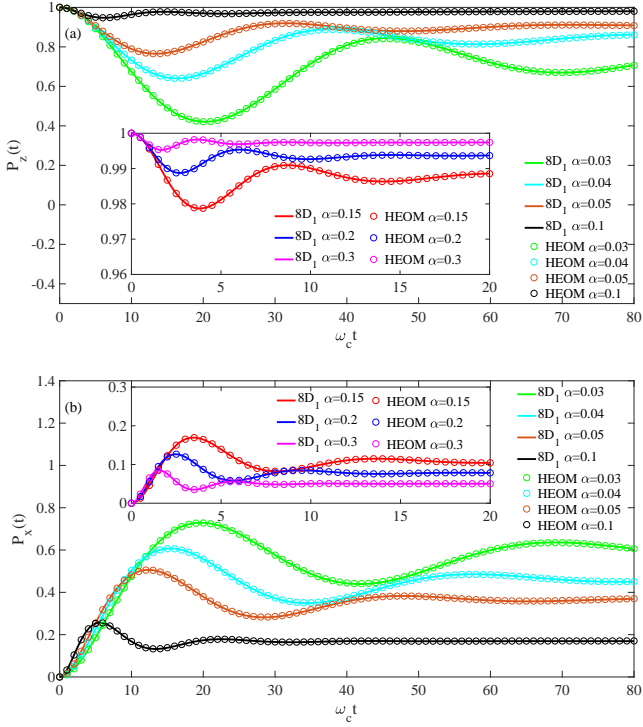}
\caption{Dynamics of $P_z(t)$ (a) and $P_x(t)$ (b) calculated by the multi-$\mathrm{D_1}$ ansatz with $M=8$ and the HEOM method for various coupling strengths $\alpha=0.03,0.04,0.05,0.1,0.15,0.2,0.3$ under the polarized bath initial condition ($\mu=1$). Other parameters are $s=0.25$, $\Delta/\omega_c=0.1$, $\epsilon=0$, $\mathrm{N_b}=200$, and $T=0$.
}
\label{fig:s025u1}
\end{figure}

\subsection{Dynamic crossover and the phase diagram}
In this section, we study the nonequilibrium quantum dynamics of the (sub-)Ohmic SBM over entire parameter regimes of spectral exponent $s$ and coupling strength $
\alpha$ under various bath initial conditions. We also identify the phase diagram of the polarization dynamics.  

\begin{figure*}[]
\centering
\includegraphics[width=17cm]{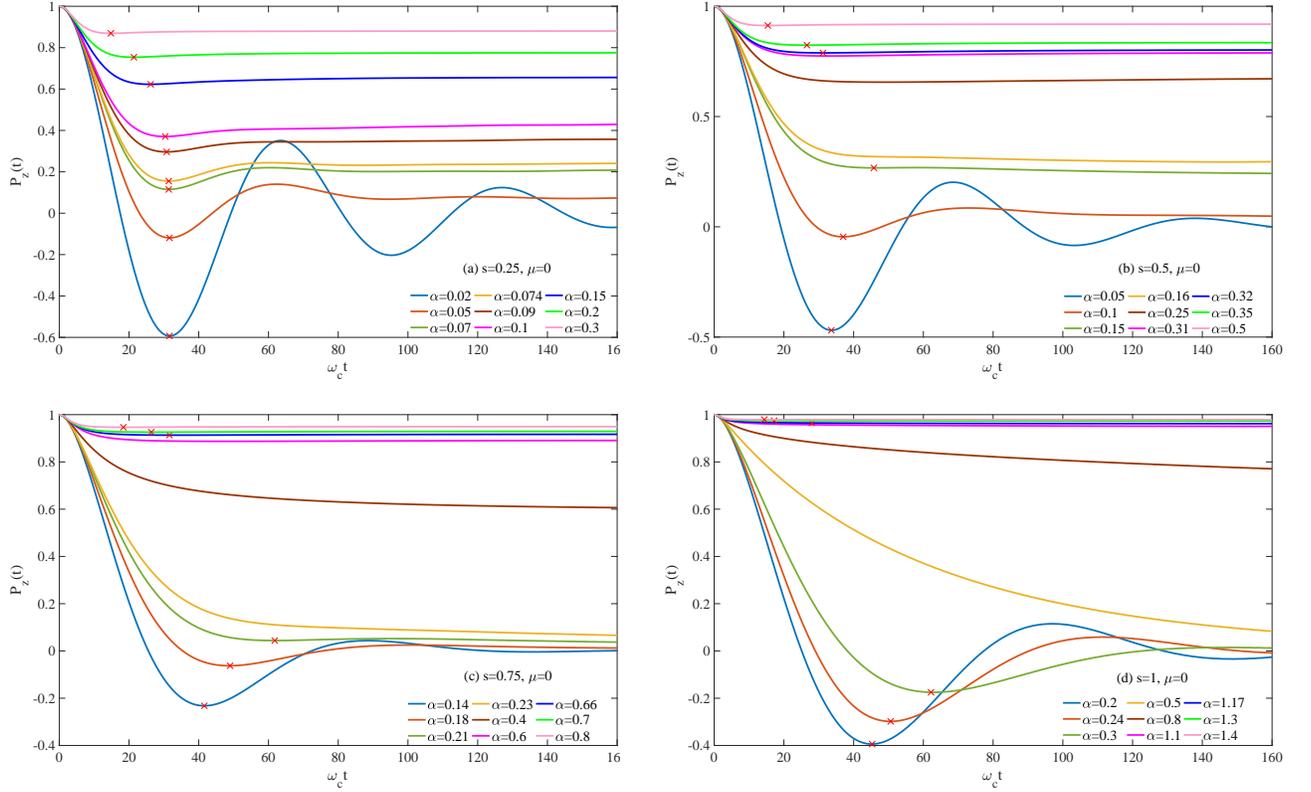}
\caption{Dynamics of $P_z(t)$ obtained by the multi-$\mathrm{D_1}$ ansatz for spectral exponents (a) $s=0.25$, (b) $s=0.5$, (c) $s=0.75$, and (d) $s=1.0$ under the factorized bath initial condition ($\mu=0$). Other parameters are $\Delta/\omega_c=0.1$, $\epsilon=0$, $\mathrm{N_b}=200$, and $T=0$. The red cross denotes the first local minimum of $P_z(t)$.}
\label{fig:Figu0}
\end{figure*}

Figure~\ref{fig:Figu0} displays the dynamics of $P_z(t)$ for spectral exponents (a) $s=0.25$, (b) $s=0.5$, (c) $s=0.75$, and (d) $s=1.0$ under the factorized bath initial condition ($\mu=0$). Let us first consider the deep sub-Ohmic case of $s=0.25$ as shown in Fig.~\ref{fig:Figu0}(a). With the increase of the coupling strength $\alpha$, one can clearly observe the persistence of coherent dynamics at coupling strengths far beyond the critical coupling of 0.022. To distinguish the coherent dynamics in weak and strong coupling regimes, we denote the dynamics as quasicoherent if $\Delta\mathrm{t_{min}}<\pi$, where $\mathrm{t_{min}}$ is the time of the first local minimum (marked by a red cross). In contrast to the coherent dynamics which is inherent to the system, the quasicoherent dynamics is purely driven by the bath. It is found that the transition from coherent to quasicoherent dynamics occurs at $\alpha=0.074$ for $s=0.25$. For $s=0.5$, the first local minimum $\mathrm{t_{min}}$ shifts to longer times as the coupling strength increases, indicating the coherent-incoherent crossover. The system exhibits purely incoherent dynamics in the range of  $0.16\leq\alpha<0.32$. For larger couplings $\alpha\geq{0.32}$, the system enters the quasicoherent phase again. For the spectral exponents of $s=0.75$ and $s=1.0$, we obtain similar three dynamical phases: coherent dynamics at weak couplings, purely incoherent dynamics at intermediate couplings, and quasicoherent dynamics at larger couplings. The domain of the incoherent phase expands when the spectral exponent $s$ increases. 

\begin{figure*}[]
\centering
\includegraphics[width=17cm]{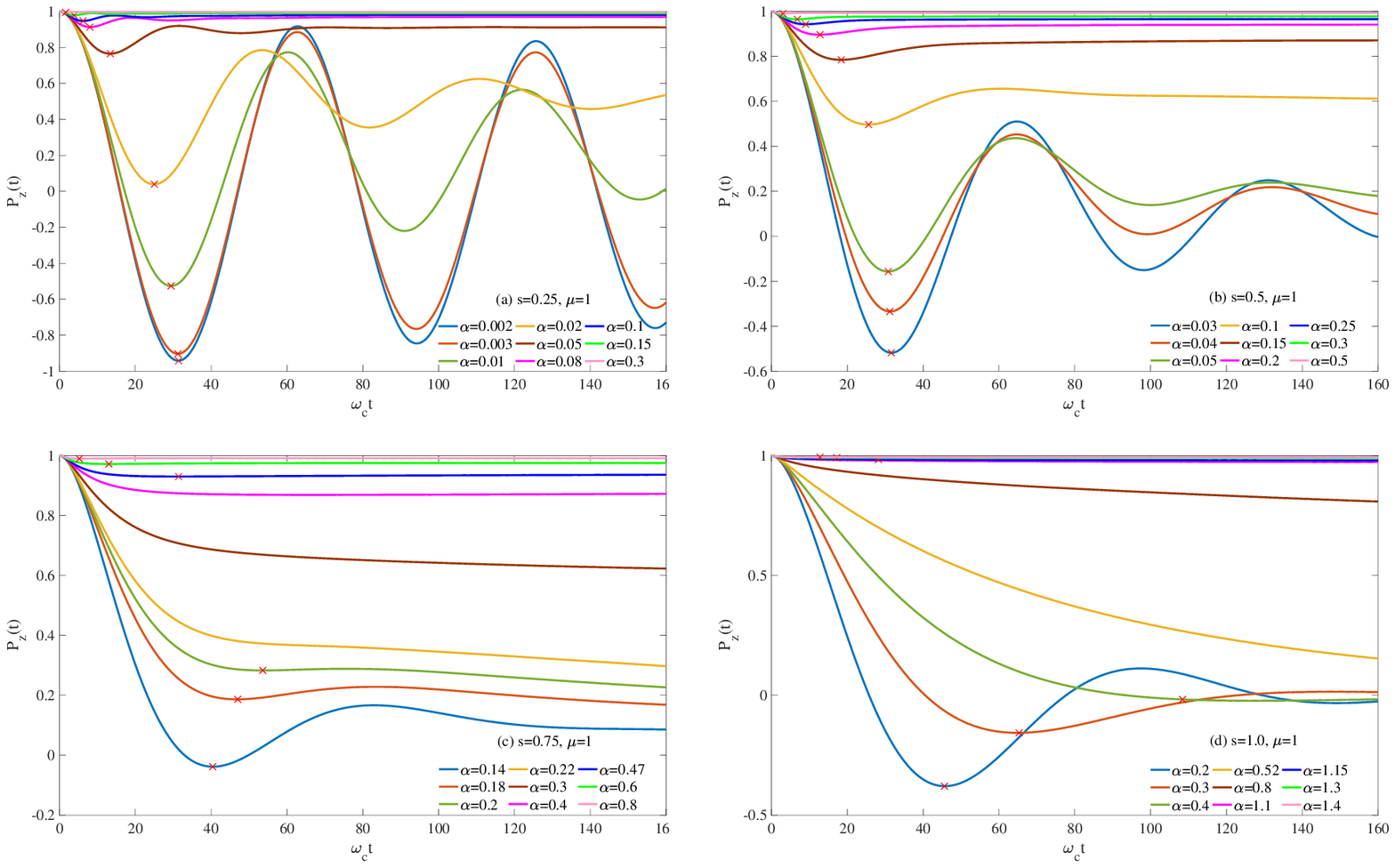}
\caption{Dynamics of $P_z(t)$ obtained by the multi-$\mathrm{D_1}$ ansatz for spectral exponents (a) $s=0.25$, (b) $s=0.5$, (c) $s=0.75$, and (d) $s=1.0$ under the polarized bath initial condition ($\mu=1$). Other parameters are $\Delta/\omega_c=0.1$, $\epsilon=0$, $\mathrm{N_b}=200$, and $T=0$. The red cross denotes the first local minimum of $P_z(t)$.
}
\label{fig:Figu1}
\end{figure*}

Figure~\ref{fig:Figu1} shows the dynamics of $P_z(t)$ for the polarized bath initial condition ($\mu=1$). For $s=0.25$, the transition from coherent to quasicoherent dynamics shifts to a much smaller coupling strength of $\alpha=0.003$ as compared to the  factorized bath initial condition ($\mu=0$). It is interesting to note that we do not find incoherent dynamics irrespective of the coupling strength $\alpha$ for $s=0.5$, quite different from the case of $\mu=0$. For $s=0.75$, we again observe transitions between three dynamical phases: coherent, incoherent and quasicoherent dynamics. The transition from incoherent to quasicoherent dynamics is found to shift to a smaller coupling strength of $\alpha=0.47$ when compared to the case of $\mu=0$. Finally, it should be mentioned that the influence of different bath initial conditions on the dynamics of $P_z(t)$ becomes weaker as the spectral exponent $s$ increases to 1 (compare Fig.~\ref{fig:Figu1}(d) and Fig.~\ref{fig:Figu0}(d)).

\begin{figure*}[]
\centering
\includegraphics[width=17cm]{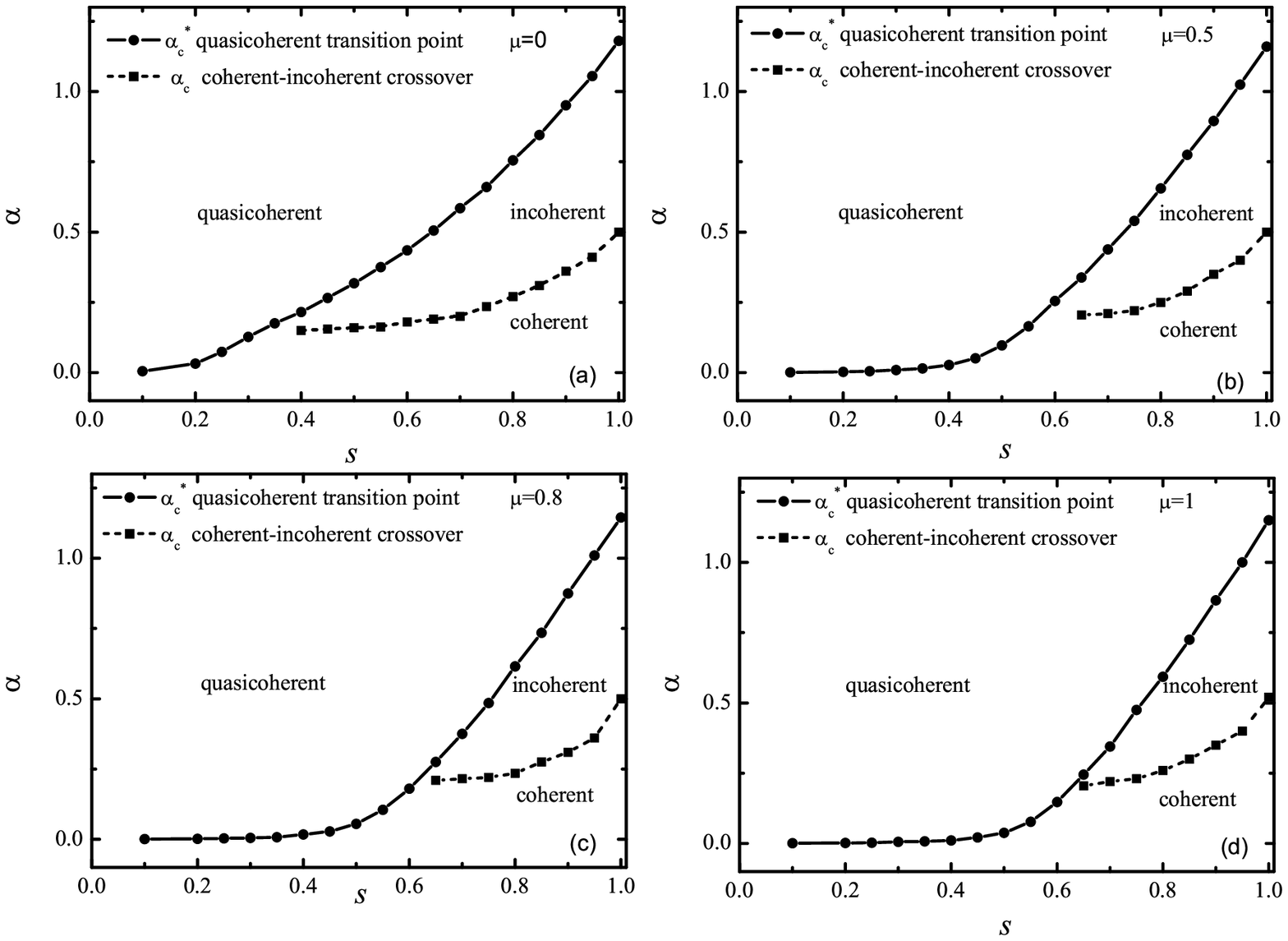}
\caption{Phase diagram of the (sub-)Ohmic SBM model at $T=0$ for different initial bath preparation parameters, (a) $\mu=0$, (b) $\mu=0.5$, (c) $\mu=0.8$, and (d) $\mu=1$. Other parameters are $\Delta/\omega_c=0.1$, $\epsilon=0$. 
}
\label{fig:FigPhase}
\end{figure*}

The phase diagrams for different bath initial conditions are sketched in Fig.~\ref{fig:FigPhase}. The general picture is that there exists a critical exponent $s_c$. For $s_c<s\leq{1}$, one can observe the transitions between three dynamical phases: the coherent, incoherent and quasicoherent dynamics. The coherent to incoherent crossover occurs at a certain coupling strength $\alpha$ (marked as black squares). Increasing $\alpha$ to a much larger value, we enter the quasicoherent phase (marked as black circles). For $s<s_c$, only the coherent to quasicoherent transition exists. As shown in Fig.~\ref{fig:FigPhase}, the initial preparation of the bath has considerable influence on the phase diagram. We note that the domain of the quasicoherent phase expands, while that of the incoherent phase shrinks when $\mu$ increases from 0 to 1. The estimated critical exponents $s_c$ for the cases of $\mu=0$, 0.5, 0.8, 1 are 0.38, 0.58, 0.62 and 0.65, respectively.  

\section{Conclusions}
In summary, we have proposed a novel description of nonequilibrium dynamics of the (sub-)Ohmic SBM under different bath initial conditions by extending the Dirac-Frenkel time-dependent variational approach with the multiple Davydov $\mathrm{D_1}$ ansatz in the interaction picture. Our approach has been validated by comparing the dynamics of $P_z(t)$ and $P_x(t)$ calculated by the multi-$\mathrm{D_1}$ ansatz with those from the HEOM method, demonstrating that our approach is capable of accurately treating dynamics of the (sub-)Ohmic SBM over a broad range of parameter regimes. By analyzing the dynamics of $P_z(t)$, we identify the phase diagram of the (sub-)Ohmic SBM. It is found that there exists a critical exponent $s_c$. For $s<s_c$, there is a transition from coherent to quasicoherent dynamics with increasing $\alpha$. For $s_c<s\leq{1}$, the system first enters the incoherent phase as $\alpha$ increases, and the quasicoherent phase emerges at a much larger value of $\alpha$. Furthermore, the dynamics of $P_z(t)$ are strongly influenced by the initial preparation of the bath. We observe that the critical exponent $s_c$ shifts to a larger value when $\mu$ increases from 0 to 1.

Realistic systems are normally subjected to environmental thermal fluctuations, which requires taking the finite temperature effect into account. Our approach can be readily extended to treat the finite temperature effect by employing the thermo field dynamics approach. \cite{LPTFD} It is also interesting to study the competitive effect of two independent baths, a pure dephasing bath and a relaxational bath, on the nonequilibrium dynamics of a two-level system, which is highly relevant for optimizing cooling protocols of the quantum devices. \cite{NalbachJCP2019} Work in these directions is in progress.

\begin{acknowledgements}
LPC  acknowledges support from the Key Research Project of Zhejiang Lab (No. 2021PE0AC02). M. F. G. acknowledges support of Hangzhou Dianzi University through startup funding. Y. Y. Yan and Z. G. L\"u are supported by the National Natural Science Foundation of China  (Grant Nos. 12005188 and 11774226) 
\end{acknowledgements}

\section*{Data availability}
Further data that support the findings of this study are available from the corresponding author upon reasonable request.

\appendix
\section{The time dependent variational approach for the SBM}\label{AppSBMVP}
By substituting the multi-$\mathrm{D}_1$ ansatz into Eq.~(\ref{VPEOM}), one readily obtains the following equations of motion 
\begin{eqnarray}
i\langle{+}|\langle{f}_l|\dot{\mathrm{D}}_1^M(t)\rangle&=&\langle{+}|\langle{f}_l|\hat{\tilde{H}}(t)|\mathrm{D}_1^M(t)\rangle, \\
iA_l^{*}\langle{+}|\langle{f}_l|\hat{b}_q|\dot{\mathrm{D}}_1^M(t)\rangle&=&A_l^{*}\langle{+}|\langle{f}_l|\hat{b}_q\hat{\tilde{H}}(t)|\mathrm{D}_1^M(t)\rangle, \\
i\langle{-}|\langle{g}_l|\dot{\mathrm{D}}_1^M(t)\rangle&=&\langle{-}|\langle{g}_l|\hat{\tilde{H}}(t)|\mathrm{D}_1^M(t)\rangle, \\
iB_l^{*}\langle{-}|\langle{g}_l|\hat{b}_q|\dot{\mathrm{D}}_1^M(t)\rangle&=&B_l^{*}\langle{-}|\langle{g}_l|\hat{b}_q\hat{\tilde{H}}(t)|\mathrm{D}_1^M(t)\rangle
\end{eqnarray}
where we have introduced 
\begin{eqnarray}
|f_l\rangle&=&\exp\left(\sum_kf_{lk}\hat{b}_k^{\dagger}-\mathrm{H.c.}\right)|\mathbf{0}\rangle \\
|g_l\rangle&=&\exp\left(\sum_kg_{lk}\hat{b}_k^{\dagger}-\mathrm{H.c.}\right)|\mathbf{0}\rangle 
\end{eqnarray}
We can further write the above equations in a matrix form
\begin{equation}
i\begin{pmatrix}
R^{(f,f)} & C^{a} &  &  \\
C^{a\dagger} & F &  &  \\
&  & R^{(g,g)} & C^b \\
& & C^{b\dagger} & G \\
\end{pmatrix}\begin{pmatrix}
\mathbf{a} \\
\mathbf{\dot{f}} \\
\mathbf{b} \\
\mathbf{\dot{g}} \\
\end{pmatrix}=\begin{pmatrix}
\mathbf{I}^{a} \\
\mathbf{I}^{f} \\
\mathbf{I}^{b} \\
\mathbf{I}^{g} \\
\end{pmatrix}
\end{equation}
where $R^{(f,f)}$ and $R^{(g,g)}$ are $M\times{M}$ matrices; $C^{a}$ and $C^{b}$ are $M\times{MN}$ matrices with $N=\mathrm{N_b}$ being the number of bath modes; $F$ and $G$ are $MN\times{MN}$ matrices; bold symbols represent the column vectors. The explicit expressions for those matrices are 
\begin{eqnarray}
R_{ln}^{(f,f)}&=&\exp\left[\sum_k\left(f_{lk}^{*}f_{nk}-\frac{1}{2}|f_{lk}|^2-\frac{1}{2}|f_{nk}|^2\right)\right], \\
R_{ln}^{(g,g)}&=&\exp\left[\sum_k\left(g_{lk}^{*}g_{nk}-\frac{1}{2}|g_{lk}|^2-\frac{1}{2}|g_{nk}|^2\right)\right], \\
C_{l,nk}^{a}&=&A_nf_{lk}^{*}R_{ln}^{(f,f)}, \\
C_{l,nk}^{b}&=&B_ng_{lk}^{*}R_{ln}^{(g,g)}, \\
F_{lq,nk}&=&A_l^{*}A_nR_{ln}^{(f,f)}(\delta_{q,k}+f_{lk}^{*}f_{nq}), \\ 
G_{lq,nk}&=&B_l^{*}B_nR_{ln}^{(g,g)}(\delta_{q,k}+g_{lk}^{*}g_{nq}).
\end{eqnarray}
The components of column vectors $\mathbf{a}=(a_1,a_2,\cdots,a_M)^T$ and $\mathbf{b}=(b_1,b_2,\cdots,b_M)^T$ are 
\begin{eqnarray}
a_n&=&\dot{A}_n-A_n\mathrm{Re}\left(\sum_k\dot{f}_{nk}f_{nk}^{*}\right), \\
b_n&=&\dot{B}_n-B_n\mathrm{Re}\left(\sum_k\dot{g}_{nk}g_{nk}^{*}\right).
\end{eqnarray}
$\mathbf{\dot{f}}$ ($\mathbf{\dot{g}}$) is an $MN\times{1}$ column vector consisting of  $\dot{f}_{nk}$ ($\dot{g}_{nk}$). $\mathbf{I}^{a}$ and $\mathbf{I}^{b}$ are $M\times{1}$ vectors; $\mathbf{I}^{f}$ and $\mathbf{I}^{g}$ are $MN\times{1}$ vectors, and the components of which are 
\begin{eqnarray}
\mathbf{I}_{l}^{a}=&&\frac{\epsilon_p(t)}{2}\sum_nA_nR_{ln}^{(f,f)}-\frac{\Delta}{2}\sum_nB_nR_{ln}^{(f,g)} \nonumber \\
&&+\sum_nA_n\sum_k\frac{\lambda_k}{2}\left(f_{lk}^{*}e^{i\omega_kt}+f_{nk}e^{-i\omega_kt}\right)R_{ln}^{(f,f)} \\
\mathbf{I}_l^b=&&-\frac{\epsilon_p(t)}{2}\sum_nB_nR_{ln}^{(g,g)}-\frac{\Delta}{2}\sum_nA_nR_{ln}^{(g,f)} \nonumber \\
&&-\sum_nB_n\sum_k\frac{\lambda_k}{2}\left(g_{lk}^{*}e^{i\omega_kt}+g_{nk}e^{-i\omega_kt}\right)R_{ln}^{(g,g)} \\
\mathbf{I}_{lq}^{f}=&&\frac{\epsilon_p(t)}{2}\sum_nA_l^{*}A_nf_{nq}R_{ln}^{(f,f)}-\frac{\Delta}{2}\sum_nA_l^{*}B_ng_{nq}R_{ln}^{(f,g)}+\sum_nA_l^{*}A_n \nonumber \\
&&\sum_k\frac{\lambda_k}{2}\left[\left(\delta_{k,q}+f_{lk}^{*}f_{nq}\right)e^{i\omega_kt}+f_{nk}f_{nq}e^{-i\omega_kt}\right]R_{ln}^{(f,f)} \\
\mathbf{I}_{lq}^{g}=&&-\frac{\epsilon_p(t)}{2}\sum_nB_l^{*}B_ng_{nq}R_{ln}^{(g,g)}-\frac{\Delta}{2}\sum_nB_l^{*}A_nf_{nq}R_{ln}^{(g,f)}-\sum_nB_l^{*}B_n \nonumber \\
&&\sum_k\frac{\lambda_k}{2}\left[\left(\delta_{q,k}+g_{lk}^{*}g_{nq}\right)e^{i\omega_kt}+g_{nk}g_{nq}e^{-i\omega_kt}\right]R_{ln}^{(g,g)} 
\end{eqnarray}
where 
\begin{equation}\label{timebias}
\epsilon_p(t)=\epsilon+y(t)+y^{*}(t)
\end{equation}
and 
\begin{equation}
R_{ln}^{(f,g)}=\left[R_{nl}^{(g,f)}\right]^{*}=\exp\left[\sum_k\left(f_{lk}^{*}g_{nk}-\frac{1}{2}|f_{lk}|^2-\frac{1}{2}|g_{nk}|^2\right)\right]
\end{equation}

\section{Hierarchy equation of motion approach for the SBM}\label{HEOMSBM}
From Eq.~(\ref{rhot}), one obtains the reduced density matrix for the system as
\begin{eqnarray}
\hat{\rho}_{\mathrm{S}}(t)=&&\mathrm{Tr}_{\mathrm{B}}\Big\{\hat{\rho}(t)\Big\}\nonumber \\
=&&\mathrm{Tr}_{\mathrm{B}}\Big\{e^{-i\hat{H}_{\mathrm{B}}t}\hat{\tilde{U}}_I(t)\hat{\rho}_{\mathrm{S}}(0)|\mathbf{0}\rangle\langle\mathbf{0}|\hat{\tilde{U}}_I^{\dagger}(t)e^{i\hat{H}_{\mathrm{B}}t}\Big\} \nonumber \\
\end{eqnarray}
where the time evolution operator $\hat{\tilde{U}}_I(t)=\hat{\mathrm{D}}^{\dagger}(\mathbf{y})\hat{U}_I(t)\hat{\mathrm{D}}(\mathbf{y})$ (see Eq.~(\ref{TTE})) corresponds to the Hamiltonian $\hat{\tilde{H}}(t)$ of Eq.~(\ref{THt}). Switching back to the Schr\"{o}dinger picture yields 
\begin{equation}
\hat{\rho}_{\mathrm{S}}(t)=\mathrm{Tr}_{\mathrm{B}}\Big\{\hat{\bar{U}}(t)\hat{\rho}_{\mathrm{S}}(0)|\mathbf{0}\rangle\langle\mathbf{0}|\hat{\bar{U}}^{\dagger}(t)\Big\}
\end{equation}
Here, the time evolution operator $\hat{\bar{U}}(t)$ corresponds to the time-dependent Hamiltonian $\hat{\bar{H}}(t)$
\begin{equation}\label{SBMTDH}
\hat{\bar{H}}(t)=\hat{H}_{\mathrm{S}}^p(t)+\sum_k\omega_k\hat{b}_k^{\dagger}\hat{b}_k+\frac{\hat{\sigma}_z}{2}\sum_k\lambda_k(\hat{b}_k^{\dagger}+\hat{b}_k)
\end{equation}
with 
\begin{eqnarray}
\hat{H}_{\mathrm{S}}^p(t)=&&\hat{H}_{\mathrm{S}}+\frac{\hat{\sigma}_z}{2}\left[y(t)+y^{*}(t)\right]\nonumber \\
=&&\frac{\epsilon_p(t)}{2}\hat{\sigma}_z-\frac{\Delta}{2}\hat{\sigma}_x
\end{eqnarray}

For the Hamiltonian of Eq.~(\ref{SBMTDH}), the reduced density matrix element for the two level system, $\hat{\rho}_{\mathrm{S}}(\sigma,\sigma^{'};t)$ ($\sigma$ is the eigenstate of the $\hat{\sigma}_z$) can be written in the path integral form as \cite{HEOMApp1,HEOMApp2,HEOMApp3}
\begin{eqnarray}
\hat{\rho}_{\mathrm{S}}(\sigma,\sigma^{'};t)=&&\int\mathcal{D}\sigma\int\mathcal{D}\sigma^{'}\hat{\rho}_{\mathrm{S}}(\sigma_0,\sigma_0^{'};t_0)\nonumber \\
&&\times{e}^{i\Theta[\sigma;t]}\Xi(\sigma,\sigma^{'};t)e^{-i\Theta[\sigma^{'};t]}
\end{eqnarray}
where $\Theta[\sigma;t]$ is the action of the two-level system, and $\Xi[\sigma,\sigma^{'};t]$ is the Feynman-Vernon influence functional which is defined as 
\begin{eqnarray}
\Xi(\sigma,\sigma^{'};t)=&&\exp\Bigg(-\int_0^{\infty}d\omega{J}(\omega)\int_{t_0}^td\tau\int_{t_0}^{\tau}d\tau^{'}\nonumber \\
&&\times\hat{V}^{\times}(\tau)\times\Big[\hat{V}^{\times}(\tau^{'})\coth(\frac{\beta\omega}{2})\cos(\omega(\tau-\tau^{'}))\nonumber \\
&&-i\hat{V}^{\circ}(\tau^{'})\sin(\omega(\tau-\tau^{'}))\Big]\Bigg)
\end{eqnarray}
where we have introduced the notations 
\begin{eqnarray}
\hat{V}=&&\frac{\hat{\sigma}_z}{2}, \nonumber \\
\hat{V}^{\times}(\tau)=&&\hat{V}[\tau]-\hat{V}[\tau^{'}], \nonumber \\
\hat{V}^{\circ}(\tau)=&&\hat{V}[\tau]+\hat{V}[\tau^{'}].
\end{eqnarray}
The bath correlation function is defined as \begin{equation}
\zeta(t)=\int_0^{\infty}d\omega{J}(\omega)\left[\coth\left(\frac{\beta\omega}{2}\right)\cos\left(\omega{t}\right)-i\sin\left(\omega{t}\right)\right]
\end{equation}
For the bath spectral density of Eq.~(\ref{SD}), $\zeta(t)$ at zero temperature can be readily obtained as 
\begin{equation}
\zeta(t)=\int_0^{\infty}J(\omega)e^{-i\omega{t}}d\omega=2\alpha\omega_c^2\Gamma(s+1)(1+i\omega_ct)^{-1-s}
\end{equation}
To facilitate the construction of HEOM, we fit $\zeta(t)$ by a sum of exponentials 
\begin{equation}
\zeta(t)=\sum_{j=1}^Jr_je^{-(\gamma_j+i\Omega_j)t}.
\end{equation}
The Feynman-Vernon influence functional can be rewritten as 
\begin{eqnarray}
\Xi(\sigma,\sigma^{'};t)=&&\exp\Bigg[-\int_{t_0}^td\tau\int_{t_0}^{\tau}d\tau^{'}\hat{V}^{\times}(\tau)\sum_j^J\Big(r_je^{-(\gamma_j+i\Omega_j)(\tau-\tau^{'})}\nonumber \\
&&\quad\quad-r_j^{*}e^{-(\gamma_j-i\Omega_j)(\tau-\tau^{'})}\Big)\hat{V}(\tau^{'})\Bigg]
\end{eqnarray}
We then introduce the auxiliary operator $\hat{\rho}_{\mathrm{S};m_{1\pm},\cdots,m_{J\pm}}(\sigma,\sigma^{'};t)$ by its matrix element as 
\begin{eqnarray}
&&\hat{\rho}_{\mathrm{S};m_{1\pm},\cdots,m_{J\pm}}(\sigma,\sigma^{'};t)\nonumber \\
=&&\int\mathcal{D}\sigma\int\mathcal{D}\sigma^{'}\hat{\rho}_{\mathrm{S}}(\sigma_0,\sigma_0^{'};t_0)\prod_{j=1}^J\Bigg(\int_{t_0}^tdse^{-(\gamma_j+i\Omega_j)(t-s)}r_j\hat{V}(s)\Bigg)^{m_{j+}}\nonumber \\
&&\Bigg(-e^{-(\gamma_j-i\Omega_j)(t-s)}r_j^{*}\hat{V}(s)\Bigg)^{m_{j-}}{e}^{i\Theta[\sigma;t]}\Xi(\sigma,\sigma^{'};t)e^{-i\Theta[\sigma^{'};t]}
\end{eqnarray}

Differentiating $\hat{\rho}_{\mathrm{S};m_{1\pm},\cdots,m_{J\pm}}(\sigma,\sigma^{'};t)$ with respect to $t$, one obtains the following HEOM
\begin{eqnarray}
&&\frac{\partial}{\partial{t}}\hat{\rho}_{\mathrm{S};m_{1\pm},\cdots,m_{J\pm}}(t) \nonumber \\
=&&-i[\hat{H}_{\mathrm{S}}^p(t),\hat{\rho}_{\mathrm{S};m_{1\pm},\cdots,m_{J\pm}}(t)]\nonumber \\
&&-\sum_j^J\Big[m_{j+}(\gamma_j+i\Omega_j)+m_{j-}(\gamma_j-i\Omega_j)\Big]\hat{\rho}_{\mathrm{S};m_{1\pm},\cdots,m_{J\pm}}(t)\nonumber \\
&&-\hat{V}^{\times}\sum_j^J\Big(\hat{\rho}_{\mathrm{S};m_{1\pm},\cdots,m_{j+}+1,m_{j-},\cdots,m_{J\pm}}(t)+\nonumber \\
&&\quad\quad\quad\quad\hat{\rho}_{\mathrm{S};m_{1\pm},\cdots,m_{j+},m_{j-}+1,\cdots,m_{J\pm}}(t)\Big)\nonumber \\
&&+\sum_j^Jm_{j+}r_j\hat{V}\hat{\rho}_{\mathrm{S};m_{1\pm},\cdots,m_{j+}-1,m_{j-},\cdots,m_{J\pm}}(t)\nonumber \\
&&-\sum_j^Jm_{j-}r_j^{*}\hat{\rho}_{\mathrm{S};m_{1\pm},\cdots,m_{j+},m_{j-}-1,\cdots,m_{J\pm}}(t)\hat{V}
\end{eqnarray}
Here $\hat{\rho}_{\mathrm{S};0,\cdots,0}(t)=\hat{\rho}_{\mathrm{S}}(t)$ represents the true reduced density matrix, while other auxiliary density matrices account for the non-perturbative and non-Markovian effects. HEOM consists of infinite set of coupled equations, which must be truncated at a finite number of hierarchy elements. In this work, we use a limited simplex truncation scheme, i.e., the integers $m_{j\pm}$ should satisfy 
\begin{equation}
\sum_j^Jm_{j\pm}<\mathrm{N_{trun}},\quad{m}_{j\pm}<=M_{j\pm}(j=1,\cdots,J) 
\end{equation}
where $\mathrm{N_{trun}}$ is the depth of the hierarchy, and $M_{j\pm}$ is the upper bound for each $m_{j\pm}$ ($j=1,\cdots,J$)


\end{document}